\documentclass{moriond}

\usepackage{natbib}
\usepackage{amsmath}
\usepackage{amssymb}
\usepackage{psfrag,epsfig,graphicx,graphics}
\usepackage{caption}
\setcitestyle{numbers}
\setcitestyle{square}
\setcitestyle{comma}

\def\meson{m}

\def\be{\begin{equation}}
\def\bea{\begin{eqnarray}}
\def\eea{\end{eqnarray}}

\newcommand{\Photo}{}

\usepackage[utf8]{inputenc}
\def\lsim{\raise0.3ex\hbox{$<$\kern-0.75em\raise-1.1ex\hbox{$\sim$}}}
\def\gsim{\raise0.3ex\hbox{$>$\kern-0.75em\raise-1.1ex\hbox{$\sim$}}}

\def\bei{\begin{itemize}}
\def\ei{\end{itemize}}
\def\bea{\begin{eqnarray}}
\def\eea{\end{eqnarray}}
\def\beas{\begin{eqnarray*}}
\def\eeas{\end{eqnarray*}}
\def\beqas{\begin{eqnarray*}}
\def\eqas{\end{eqnarray*}}
\def\beq{\begin{equation}}
\def\eeq{\end{equation}}
\def\beqd{\begin{displaymath}}
\def\eeqd{\end{displaymath}}
\def\eqd{\end{displaymath}}

\def\beeq{\begin{eqnarray}} \def\eeeq{\end{eqnarray}}

\def\bef{\begin{frame}}

\def\slashchar#1{\setbox0=\hbox{$#1$}
\dimen0=\wd0
\setbox1=\hbox{/} \dimen1=\wd1
\ifdim\dimen0>\dimen1
\rlap{\hbox to \dimen0{\hfil/\hfil}}
#1
\else
\rlap{\hbox to \dimen1{\hfil$#1$\hfil}}
/
\fi}

\newcommand{\ee}{\end{equation}}
\newcommand{\eq}{\end{equation}}

\newcommand{\fin}{\end{document}}

\def\pv{\vec{p}_t}
\def\dv{\vec{\Delta}_t}

\def\beqa{\begin{eqnarray}}
\def\eqa{\end{eqnarray}}

\begin{document}
\title{Accessing GPDs through the exclusive photoproduction of a $ \gamma  $-meson pair~\footnote{Presented by S.~Nabeebaccus at DIS2022: XXIX International Workshop on Deep-Inelastic Scattering and Related Subjects, Santiago de Compostela, Spain, May 2-6 2022.}}

\author{G.~Duplan\v{c}i\'{c}$ ^{1} $, S.~Nabeebaccus$ ^{2} $, K.~Passek-Kumeri\v{c}ki$ ^{1} $, B.~Pire$ ^{3} $, L. Szymanowski$ ^{4} $ and S.~Wallon$ ^{2} $}

\address{\vspace{0.4cm}$^1$Theoretical Physics Division, Rudjer Bo{\v s}kovi{\'c} Institute,
	HR-10002 Zagreb, Croatia\\
$^2$Universit\'e Paris-Saclay, CNRS/IN2P3, IJCLab, 91405 Orsay, France\\  
$^3$Centre de Physique Th\'eorique, CNRS, Ecole polytechnique, I.P. Paris, 91128 Palaiseau, France\\  
$^4$National Centre for Nuclear Research (NCBJ), Warsaw, Poland 
}

\maketitle

\abstracts{\vspace{0.1cm}We consider the exclusive photo-production of a photon-meson pair with a large invariant mass, working in the QCD factorisation framework. Explicitly, we consider a $ \rho $-meson or a charged $  \pi  $ in the final state. This process gives access both to chiral-even GPDs and chiral-odd GPDs, which are not well-known experimentally, especially the latter ones. The computation is performed at leading order and leading twist. We discuss the prospects of measuring them in experiments, focusing on the kinematics at the JLab 12-GeV experiment, and pPb ultra-peripheral collisions at LHC. In particular, the latter gives access to the small $  \xi $ regime of GPDs.}

\section{Introduction}

\begin{figure}[h]
	
	\psfrag{TH}{$\Large T_H$}
	\psfrag{Pi}{$m'$}
	\psfrag{P1}{$\,\phi$}
	\psfrag{P2}{$\,\phi$}
	\psfrag{Phi}{$\,\phi$}
	\psfrag{Rho}{$ m $}
	\psfrag{tp}{$t'$}
	\psfrag{s}{$s$}
	\psfrag{x1}{$\!\!\!\!\!\!x+\xi$}
	\psfrag{x2}{$\!\!x-\xi$}
	\psfrag{RhoT}{$M_T$}
	\psfrag{t}{$t$}
	\psfrag{N}{$N$}
	\psfrag{Np}{$N'$}
	\psfrag{M}{$M^2_{\gamma m }$}
	\psfrag{GPD}{$\!GPD$}

	\centerline{
		\raisebox{1.6cm}{\includegraphics[width=12pc]{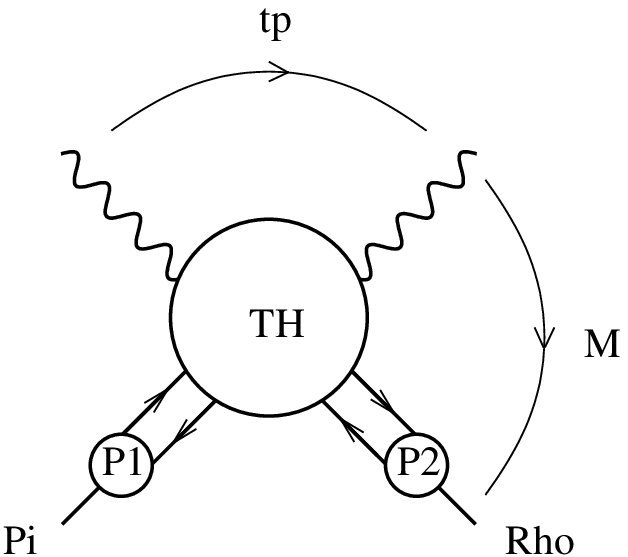}}~~~~~~~~~~~~~~
		\psfrag{TH}{$\,\Large T_H$}
			\psfrag{Rho}{$ m $}
			\psfrag{M}{$M^2_{\gamma m }$}
		\includegraphics[width=12pc]{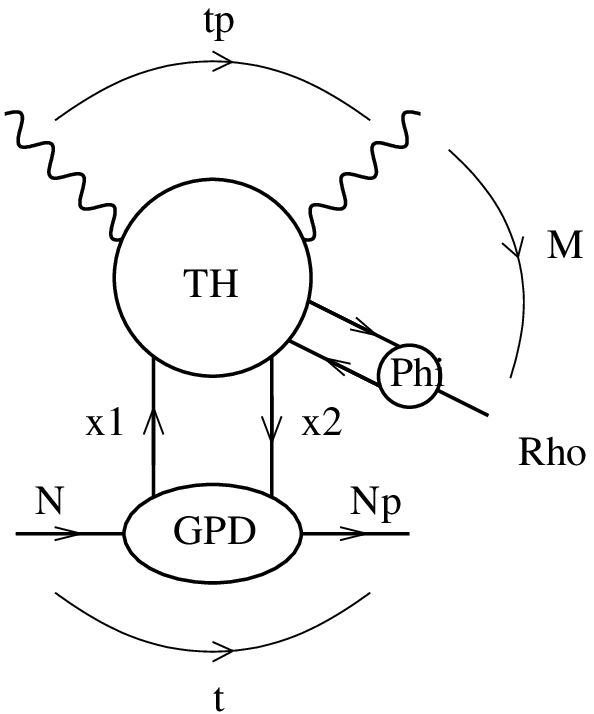}}
	
	\caption{\small Left: Factorization of the amplitude for the process $\gamma + m' \rightarrow \gamma + m  $ at large $s$ and fixed angle (i.e. fixed ratio $t'/s$). Right:  Replacing the $\pi$ meson distribution amplitude by a nucleon generalized parton distribution  leads to the factorization of the amplitude  for $\gamma + N \rightarrow \gamma + m  +N'$ at large $M_{\gamma m }^2$.}
	\label{Fig:feyndiag}
\end{figure}

Generalised parton distributions (GPDs) have been extensively studied in the context of deeply virtual Compton scattering (DVCS) and deeply virtual meson production. This allows us to extract information on the 3D structure of nucleons. Another channel was proposed to study GPDs in \cite{Boussarie:2016qop,Duplancic:2018bum} - the photoproduction of a photon-meson pair with a large invariant mass $ M_{\gamma \meson}^2 $. Imposing the latter constraint is important as it provides the hard scale for collinear QCD factorisation. Its use can be justified by comparing this process with photon-meson scattering at large angles, as illustrated in Figure \ref{Fig:feyndiag}. In fact, in the closely-related processes of diphoton photoproduction \cite{Grocholski:2022rqj} and exclusive production of a photon pair from pion-nucleon collisions \cite{Qiu:2022bpq}, where the photon pair has a large invariant mass, collinear QCD factorisation has been shown to hold at leading twist.

The amplitude for our process is thus expressed as the convolution of a \textit{hard part} (coefficient function), which is calculated using perturbative techniques, with two soft parts, namely a GPD involving the incoming and outgoing nucleons and a distribution amplitude (DA) for the outgoing meson. One of the main advantages of studying this channel is that for a transversely polarised $  \rho  $ meson, this process gives access to chiral-odd GPDs at \textit{leading twist}, unlike in deeply virtual meson production. This is interesting, as chiral-odd GPDs are not well-known experimentally.

\section{Kinematics}

The process we study is
\begin{align*}
\gamma (q)+N(p_1) \longrightarrow \gamma (k)+N'(p_2)+m(p_{\meson})\;,
\end{align*}
where $ \meson= \rho^{0,\pm}_{L,T},  \pi ^{\pm} $. Using two light-cone vectors $p$ and $n$ (with $p \cdot n = \frac{s}2$), the particle momenta can be written as
\begin{eqnarray}
\label{impini}
p_1^\mu = (1+\xi)\,p^\mu + \frac{M^2}{s(1+\xi)}\,n^\mu~, \quad p_2^\mu = (1-\xi)\,p^\mu + \frac{M^2+\vec{\Delta}^2_t}{s(1-\xi)}n^\mu + \Delta^\mu_\bot\,, \quad q^\mu = n^\mu ~,
\end{eqnarray}
\begin{eqnarray}
\label{impfinc}
k^\mu = \alpha \, n^\mu + \frac{(\vec{p}_t-\vec\Delta_t/2)^2}{\alpha s}\,p^\mu + p_\bot^\mu -\frac{\Delta^\mu_\bot}{2}~, \quad \! \!
p_\meson^\mu = \alpha_\meson \, n^\mu + \frac{(\vec{p}_t+\vec\Delta_t/2)^2+M^2_\meson}{\alpha_\meson s}\,p^\mu - p_\bot^\mu-\frac{\Delta^\mu_\bot}{2}\,,\quad \nonumber
\end{eqnarray}
where $M$ and $M_\meson$ are the masses of the nucleon and the meson respectively. The  square of the centre of mass energy of the $\gamma$-N system is
then $
S_{\gamma N} = (q + p_1)^2 = (1+\xi)s + M^2$, while the small  squared transferred momentum is
$
t = (p_2 - p_1)^2 = -\frac{1+\xi}{1-\xi}\vec{\Delta}_t^2 -\frac{4\xi^2M^2}{1-\xi^2}$. The hard scale $M^2_{\gamma\meson}$ is the invariant mass squared of the $\gamma$-meson system. This is imposed by having a large \textit{relative} transverse momentum $  \vec{p}_{t}  $ between the outgoing photon and meson.

The use of collinear QCD factorisation requires that $ -u'= \left( p_{\meson}-q \right)^2  $, $ -t'= \left( k-q \right)^2  $ and $ M_{\gamma \meson}^2 =  \left(  p_{\meson}+k\right)^2  $ to be large, while $ -t =  \left( p_2-p_1 \right)^2  $ needs to be small. For this, we employ the cuts $ -u',-t'>1$ GeV$ ^2 $, and $ -t < 0.5 $ GeV$ ^2 $. We note that these cuts are sufficient to ensure that $ M^2_{\gamma\meson} > 1 $ GeV$ ^2 $.

In the generalized Bjorken limit, neglecting  $\dv$ in front of $\pv$, as well as hadronic masses, we have that the approximate kinematics is
\begin{eqnarray}
\label{skewness2}
M^2_{\gamma\meson} \approx  \frac{\vec{p}_t^2}{\alpha\bar{\alpha}} ~, \qquad
 \alpha _{\meson} \approx 1-\alpha \equiv \bar{\alpha} ~,\qquad
\xi =  \frac{\tau}{2-\tau} ~,
\end{eqnarray}
\begin{eqnarray}
\tau \approx 
\frac{M^2_{\gamma\meson}}{S_{\gamma N}-M^2}~,\qquad
~-t'  \approx  \bar\alpha\, M_{\gamma\meson}^2  ~,\qquad -u'  \approx  \alpha\, M_{\gamma\meson}^2 \,.\quad \,\nonumber
\end{eqnarray}
In expressing the results, we will use  $ \left( -u' \right),\,(-t)$, and $ M_{\gamma \meson}^2 $  as differential variables.

\section{Non-perturbative inputs: GPDs and DAs}

The chiral-even light-cone DA, e.g. for the longitudinally polarized $\rho^0_L$-meson is defined, at the leading 
twist 2, by the matrix element \cite{Ball:1996tb},
\begin{equation}
\langle 0|\bar{u}(0)\gamma^\mu u(x)|\rho_{L}^0(p_\rho) \rangle = \frac{1}{\sqrt{2}}p_\rho^\mu f_{\rho^0}\int_0^1dz\ e^{-izp_\rho\cdot x}\ \phi_{\parallel}(z),
\label{defDArhoL}
\end{equation}
where the decay constant $ f_{ \rho ^0}= 213$ MeV. The DAs for the $  \rho _{T} $ and $  \pi^{\pm} $ mesons are defined analogously. For the computation, we use the asymptotic form of the distribution amplitude, as well as an alternative form, which we call `holographic' DA, given by
\beqa
\label{DA-asymp}
\phi^{\rm as}(z)= 6 z (1-z)\;,\quad \phi^{\rm hol}(z)= \frac{8}{ \pi } \sqrt{z (1-z)}\;,
\eqa
where both are normalised to 1. The alternative form, first proposed in \cite{Mikhailov:1986be}, has been suggested in the literature in the context of AdS-QCD holographic correspondence \cite{Brodsky:2006uqa} (hence the name `holographic' DA) and dynamical chiral symmetry breaking on the light-front \cite{Shi:2015esa}. In fact, recent lattice results indicate an even further departure from the asymptotical form, with $  \phi (z) \propto z^{ \alpha } \left( 1-z \right)^{ \alpha }  $ and $  \alpha \approx 0.2-0.32 $ \cite{Gao:2022vyh}.

We now turn to the definition of the GPDs. The chiral-even GPDs of a parton $q$ (where $q = u,\ d$) in the nucleon target   ($\lambda$ and $\lambda'$ are the light-cone helicities of the nucleons with momenta $p_1$ and $p_2$) are  defined by~\cite{Diehl:2003ny}:
\beqa
&&\langle p(p_2,\lambda')|\, \bar{q}\left(-\frac{y}{2}\right)\,\gamma^+q \left(\frac{y}{2}\right)|p(p_1,\lambda) \rangle \\ \nonumber 
&&= \int_{-1}^1dx\ e^{-\frac{i}{2}x(p_1^++p_2^+)y^-}\bar{u}(p_2,\lambda')\, \left[ \gamma^+ H^{q}(x,\xi,t)   +\frac{i}{2m}\sigma^{+ \,\alpha}\Delta_\alpha  \,E^{q}(x,\xi,t) \right]
u(p_1,\lambda)\,,
\label{defGPDEvenV}
\eqa
and analogously for chiral-even axial GPDs. In our analysis, 
the contributions from $ E^{q} $ and $  \tilde{E}^{q}  $ (in the chiral-even axial GPD) are neglected, since they are suppressed by kinematical factors at the cross-section level.

The transversity GPD of a quark $q$   is defined by:
\beqa
&&\langle p(p_2,\lambda')|\, \bar{q}\left(-\frac{y}{2}\right)i\,\sigma^{+j} q \left(\frac{y}{2}\right)|p(p_1,\lambda)\rangle \\ \nonumber
&&= \int_{-1}^1dx\ e^{-\frac{i}{2}x(p_1^++p_2^+)y^-}\bar{u}(p_2,\lambda')\, \left[i\,\sigma^{+j}H_T^{q}(x,\xi,t)
+\dots
\right]u(p_1,\lambda)\,,
\label{defGPD}
\eqa
where $\dots$ denote the remaining three chiral-odd GPDs whose contributions are omitted in our calculation for the same reasons as mentioned above. The GPDs are parametrised in terms of double distributions \cite{Radyushkin:1998es}.

We note that for the modelling of the chiral-even axial GPDs and chiral-odd GPDs, we use two different parametrisations for the input PDFs: The \textit{standard} scenario, for which the light sea quark and anti-quark distributions are \textit{flavour-symmetric}, and the \textit{valence} scenario which corresponds to completely \textit{flavour-asymmetric} light sea quark densities. More details on the modelling of the GPDs can be found in \cite{Boussarie:2016qop,Duplancic:2018bum}.

\section{Computation}

The hard part of the amplitude requires the computation of 20 Feynman diagrams in total. One can exploit the $ C $-symmetry of the process to reduce the number of diagrams by half. The convolution over $ z $ with the distribution amplitude (taken to have the two forms in \eqref{DA-asymp}) is then performed \textit{analytically}, while the convolution over $ x $ with the GPD is performed \textit{numerically}, in terms of a few building block integrals.

The fully differential cross-section, as a function of $ -u' $, $ -t $ and $ M_{\gamma \meson}^2 $,  is then given by
\begin{equation}
\label{difcrosec}
\left.\frac{d\sigma}{dt \,du' \, dM^2_{\gamma\meson}}\right|_{\ -t=(-t)_{ \mathrm{min} }} = \frac{|\mathcal{\overline{M}}|^2}{32S_{\gamma N}^2M^2_{\gamma\meson}(2\pi)^3}\;,
\end{equation}
where $ -t $ has been set to the minimum value $ (-t)_{ \mathrm{min} } $ allowed by the kinematics, including the imposed cuts, and is in general a function of $ M^2_{\gamma \meson} $ and $ S_{\gamma N} $ itself. We refer the reader to \cite{Boussarie:2016qop,Duplancic:2018bum} for the details regarding the computation.

Performing the integration over $ -u' $ and $ -t $ gives
\begin{align}
\label{difcrosec2}
\frac{d\sigma}{dM^2_{\gamma\meson}} = \int_{(-t)_{ \mathrm{min} }}^{(-t)_{ \mathrm{max} }} \ d(-t)\ \int_{(-u')_{ \mathrm{min} }}^{(-u')_{ \mathrm{max} }} \ d(-u') \ F^2_H(t)\times\left.\frac{d\sigma}{dt \, du' d M^2_{\gamma\meson}}\right|_{\ -t=(-t)_{min}} \;,
\end{align}
where $ F_H(t)=\frac{ \left( t_{ \mathrm{min}}-C \right)^2  }{\left( t-C \right)^2} $ is a standard dipole form factor, with $ C=0.71 $ GeV$ ^2 $.

\section{Results}

\begin{figure}[h]
	\vspace{0.3cm}

	\psfrag{HH}{\hspace{-1.5cm}\raisebox{-.6cm}{\scalebox{.8}{$-u' ({\rm 
					GeV}^{2})$}}}
		\psfrag{VVV}{\raisebox{.3cm}{\scalebox{.8}{$\hspace{-.4cm}\displaystyle\left.\frac{d 
						\sigma_{\gamma\rho_{L}^0}}{d M^2_{\gamma \rho} d(-u') d(-t)}\right|_{(-t)_{\rm min}}({\rm pb} \cdot {\rm GeV}^{-6})$}}}
	\psfrag{TT}{}
	
	\psfrag{HHHM2}{\hspace{-1.5cm}\raisebox{-.6cm}{\scalebox{.8}{$M^{2}_{\gamma  \rho } ({\rm 
					GeV}^{2})$}}}
	\psfrag{VVVM2}{\raisebox{.3cm}{\scalebox{.8}{$\hspace{-.4cm}\displaystyle\frac{d 
					\sigma_{\gamma\rho_{L}^0}}{d M^2_{\gamma \rho} }({\rm pb} \cdot {\rm GeV}^{-2})$}}}
	\psfrag{TTTM2}{}

	\centerline{	\hspace{0.35cm}	{\includegraphics[width=19pc]{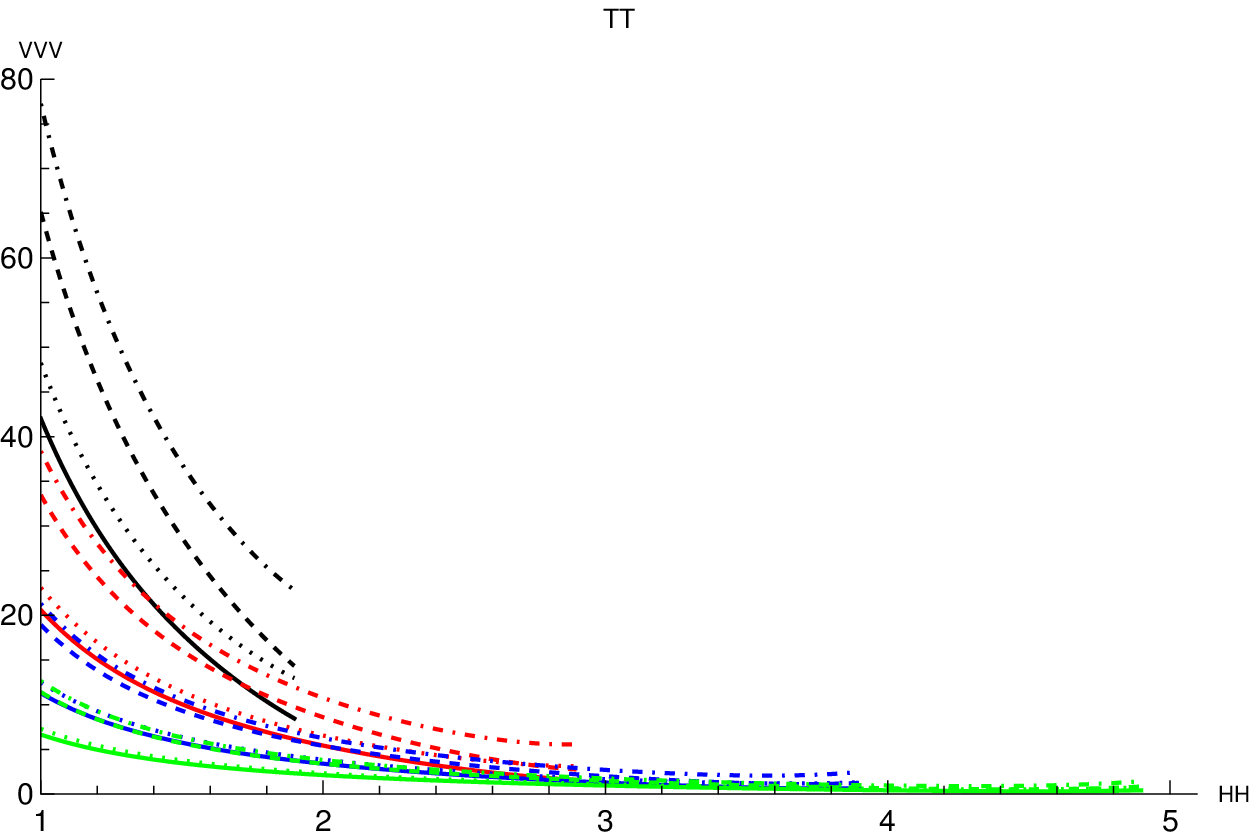}}
\includegraphics[width=19pc]{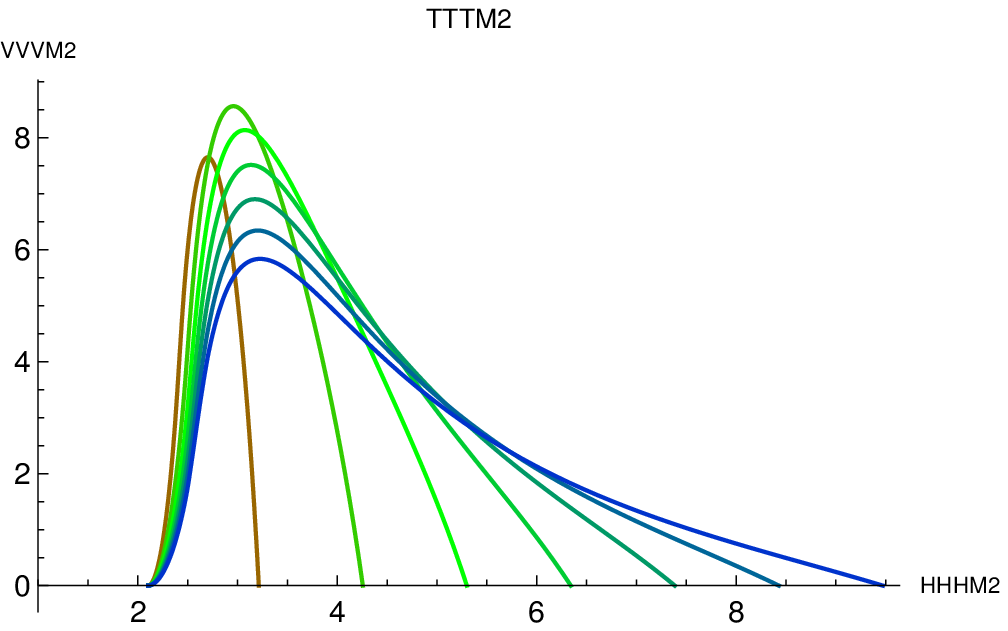}}

	\vspace{0.4cm}

		\caption{\small Left: The fully differential cross-section for $  \rho _{L}^0 $ as a function of $ -u' $ is shown. $ M_{\gamma  \rho }^2 =3,4,5,6$ GeV$ ^2 $ correspond to black, red, blue and green respectively. The difference between standard (dotted) and valence (solid) scenarios for an asymptotical DA, and between standard (dot-dashed) and valence (dashed) scenarios for a holographic DA is also illustrated.  Here, $ S_{\gamma N}=20 $ GeV$ ^2 $. Right: The single differential cross-section for $  \rho _L^0 $ as a function of $ M_{\gamma  \rho }^2 $ using a holographic DA and the standard scenario. The values of $ S_{\gamma N} $ used are 8 (brown), 10, 12, 14, 16, 18, 20 (blue) from left to right.}
	\label{Fig:diffXsection}
\end{figure}

Due to lack of space, we present only a few plots which are representative, focusing on the longitudinally polarised $  \rho _{L}^0 $ meson on a proton target. In Figure \ref{Fig:diffXsection} on the left, we show the fully differential rate as a function of $ -u' $, for different values of $ M_{\gamma \meson}^2 $. The effect of using the two different models for the distribution amplitude, as well as that of using the valence and standard scenarios for modelling the GPDs, is also illustrated. We thus find that using the holographic DA gives a result that is roughly twice that of the asymptotical DA. Nevertheless, to properly distinguish between the two models, one would need to include NLO corrections, since they can be large.

The single differential cross-section as a function of $ M_{\gamma  \rho } ^2$ for different values of $ S_{\gamma N} $ is shown on the right plot in Figure \ref{Fig:diffXsection}. Only the result for the holographic DA with the standard scenario is shown, since it gives the largest contribution, as illustrated in the plot of the fully differential cross-section. We note that while the fully differential cross-section is largest for smaller $ M_{\gamma  \rho }^2 $, the range of $ -u' $ is more restricted, due to the shrinking of the phase space. In fact, there is a compromise between the two effects, and this explains the position of the peak around $ M_{\gamma  \rho }^2 \approx 3 $ GeV$ ^2 $ in the single differential cross-section plot on the right of Figure \ref{Fig:diffXsection}. We also note that the position of this peak is more or less the same for different $ S_{\gamma N} \leq 2000 $ GeV$ ^2 $.

\begin{figure}[h!]
	
	\psfrag{HHH}{\hspace{-1.5cm}\raisebox{-.6cm}{\scalebox{.8}{$S_{\gamma N} ({\rm 
					GeV}^{2})$}}}
	\psfrag{VVV}{\raisebox{.3cm}{\scalebox{.9}{$\hspace{-.4cm}\displaystyle
					\sigma_{\gamma\rho_{L}^0}({\rm pb} )$}}}
	\psfrag{TTT}{}
	
	\psfrag{HHHp}{\hspace{-1.5cm}\raisebox{-.6cm}{\scalebox{.8}{$S_{\gamma N} ({\rm 
					GeV}^{2})$}}}
	\psfrag{VVVp}{\raisebox{.3cm}{\scalebox{.9}{$\hspace{-.4cm}\displaystyle
				\frac{d N _{\gamma }}{dS_{ \gamma N}}({\rm GeV^{-2}} )$}}}
	\psfrag{TTTp}{}

	\centerline{
		{\includegraphics[width=18pc]{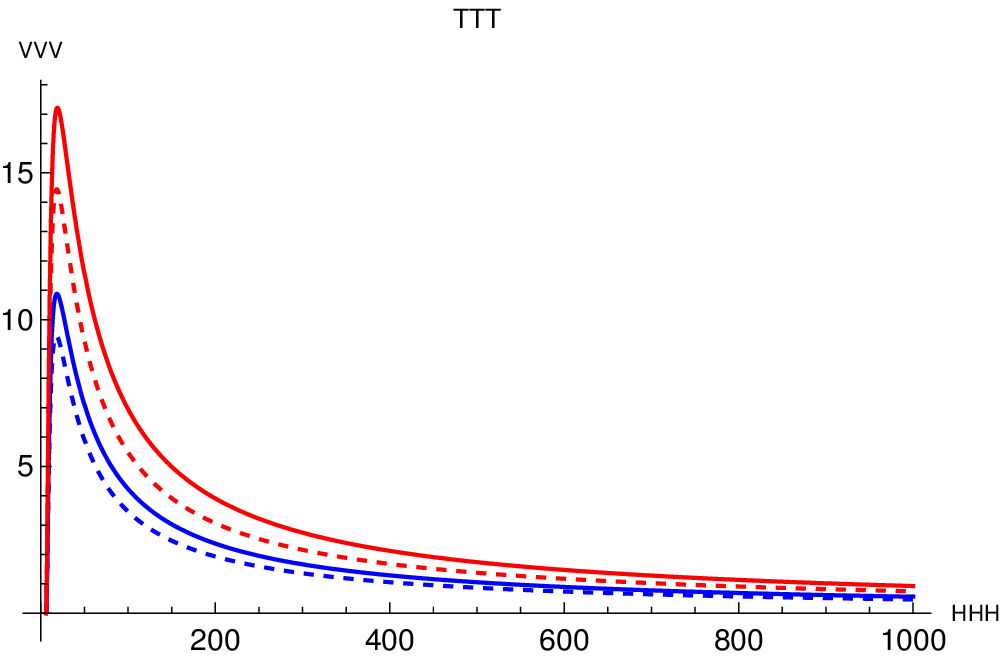}}
	{\includegraphics[width=18pc]{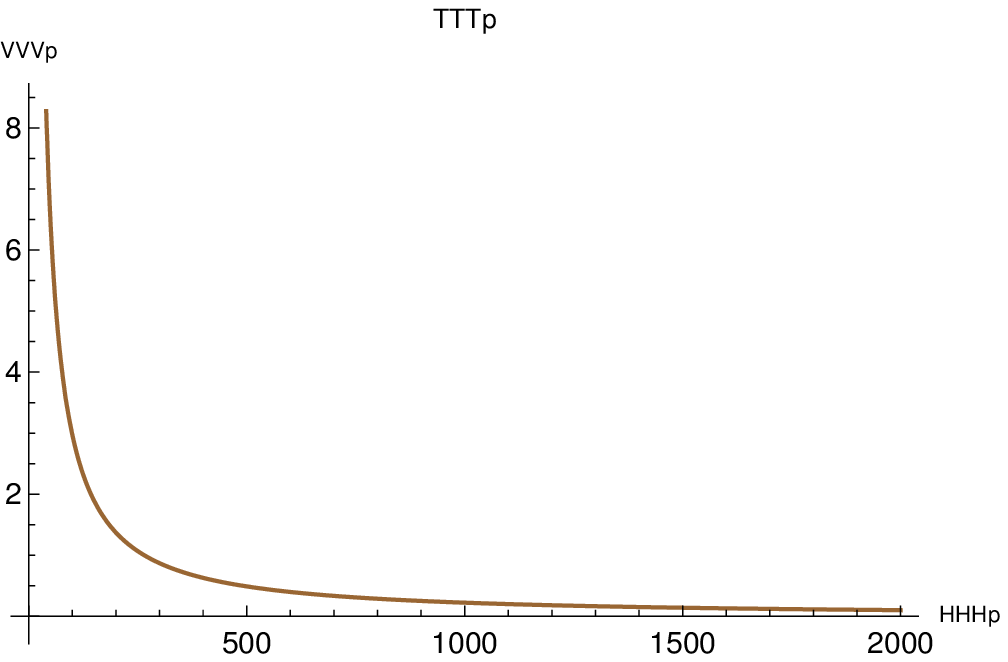}}}

\vspace{0.2cm}
	\caption{\small Left: The plot shows the cross-section $  \sigma _{\gamma  \rho _L^0} $ as a function of the centre of mass energy $ S_{\gamma N} $. The holographic DA and asymptotical DA cases are shown in red and blue respectively. The standard scenario is represented by solid lines, while the valence scenario is represented by dashed lines. Right: The photon flux $ \frac{d N _{\gamma }}{dS_{ \gamma N}}$ is shown as a function of $ S_{ \gamma N} $.}
	\label{Fig:SigmavsSgN}
\end{figure}

Finally, in Figure \ref{Fig:SigmavsSgN}, we show the variation of the cross-section as a function of $ S_{\gamma N} $ (left), as well as the photon flux $ \frac{dN_{\gamma }}{dS_{\gamma  N}} $ in pPb ultra-peripheral collisions (UPCs) at LHC (right). The cross-section drops rather rapidly with $ S_{\gamma N} $, and has a peak at around 20 GeV$ ^2 $. The total cross-section for the exclusive photoproduction of a $ \gamma  \rho ^0_{L} $ pair (with large invariant mass) on a proton target in pPb collisions can be obtained simply by convoluting the photon flux with $  \sigma _{ \gamma \rho _L^0} $ (see Figure \ref{Fig:SigmavsSgN}). We note that while LHC can access very high energies, the photon flux from the Pb nucleus in pPb collisions decreases very rapidly with $ S_{\gamma N} $. This implies that the total cross-section is dominated by the region of relatively small $ S_{\gamma N} $.

The counting rates for the different mesons, using the expected luminosity ($ 1.2 $ pb$ ^{-1} $) for runs 3 and 4 at ATLAS/CMS \cite{Citron:2018lsq}, are shown in Table \ref{tab:counting-rates}. The range is obtained by considering the minimum and maximum obtained from the different models (holographic DA vs asymptotical DA, and valence vs standard scenarios). Two sets of counting rates are shown, one without any cut in $ S_{\gamma N} $ and the other with a cut of $ S_{\gamma N} \geq 300 $ $ \mathrm{GeV}^2  $. Introducing a lower bound on $ S_{\gamma N} $ allows us to study GPDs in the small $  \xi  $ region, which is very interesting. At $ S_{\gamma N} =300 $ $  \mathrm{GeV}^2 $, we find that the region of $ M_{\gamma \meson}^2  $ where the cross-section is maximum (see Figure \ref{Fig:diffXsection}) corresponds to $  \xi  \approx 5  \cdot 10^{-3} $, and it goes down to $  \xi \approx 7.5 \cdot 10^{-4}$ at $ S_{\gamma N}=2000 $ $ \mathrm{GeV}^2  $. Despite the fact that the number of events is dominated by the region of $ S_{\gamma N} \leq 300 $ $ \mathrm{GeV}^2 $, we find that there is still reasonable statistics to prompt a study of the process in the small $  \xi  $ region at LHC.

\renewcommand{\arraystretch}{1.1}

\begin{table}[h]
	\centering
	\begin{tabular}{ |c |c|c|}
		\hline
		Meson & 		Without cut & $S_{\gamma N} \geq 300 $ $ \mathrm{GeV}^2  $\\
		\hline\hline
		$  \rho _L^0 $ &  8.7--16 $ \times  10^3$ & 3.6--7.2 $ \times 10^2 $ \\
		\hline
		$  \rho _L^+  $ & 4.8--10 $ \times  10^3$& 1.9--5.6 $ \times 10^2 $\\
		\hline
		$  \pi ^+ $ & 1.6--9.2 $ \times  10^3$& 1.0--3.1 $ \times 10^2 $\\
		\hline 
	\end{tabular}
	\caption{\small The counting rates for the various mesons in pPb UPCs at LHC are shown, using the expected luminosity in Runs 3 and 4 for ATLAS/CMS. The second column shows the case without any cuts in $ S_{\gamma N} $, while the third corresponds to a cut of $ S_{\gamma N} \geq 300$ $  \mathrm{GeV}^2  $, which gives access to the small $  \xi  $ region. }
	\label{tab:counting-rates}
\end{table}

The counting rates for the JLab 12-GeV experiment, which are roughly 1 order of magnitude larger than those reported here, can be found in \cite{Boussarie:2016qop,Duplancic:2018bum}. Although the numbers are lower for pPb UPCs at LHC, the energies that can be accessed are higher.

Note that some of the results are different from those in the published papers \cite{Boussarie:2016qop,Duplancic:2018bum}, due to recent improvements of our code. The details of the computation, including various other results, will appear in a forthcoming publication.

\bibliographystyle{utphys}

\bibliography{
	/home/saadn/work/projects/common/masterrefs.bib}

\end{document}